\documentclass[fleqn,10pt]{wlscirep}
\usepackage[utf8]{inputenc}
\usepackage[T1]{fontenc}
\usepackage{authblk}
\usepackage{bm}

\title{Modelling the Climate Change Debate in Italy through Information Supply and Demand}

\author[1,+]{Irene Scalco}
\author[2,$\dagger$]{Giulia Colafrancesco}
\author[3,*]{Matteo Cinelli}

\affil[1]{Department of Social Sciences and Economics, Sapienza University of Rome, P.le Aldo Moro, 5, Rome, 00185, Italy}
\affil[2]{ECCO, The italian climate change think tank, Via del Quirinale, 26, Rome, 00184, Italy}
\affil[3]{Department of Computer Science, Sapienza University of Rome, Viale Regina Elena 295, Rome, 00161, Italy}

\affil[+]{irene.scalco@uniroma1.it}
\affil[$\dagger$]{giulia.colafrancesco@eccoclimate.org}
\affil[*]{matteo.cinelli@uniroma1.it}

\begin{abstract}
Climate change is one of the most critical challenges of the twenty-first century. Public understanding of climate issues and of the goals regarding the climate transition is essential to translate awareness into concrete actions. In this context, social media platforms play a crucial role in disseminating information about climate change and climate policy. To better understand the dynamics of information circulation and the emergence of information voids we propose a model that takes into account the supply and demand of information related to the Italian climate-transition discourse. We conceptualise information supply as the production of content on Facebook, Instagram and GDELT (an online news database) while leveraging Google searches to capture information demand. Our findings highlight responsiveness and temporal coupling between supply and demand, particularly during moments of heightened public attention triggered by significant external events. These responsive interactions reveal an overall adaptive information ecosystem. However, we also observe persistent information voids which may limit public understanding and delay meaningful engagement.

\end{abstract}
\begin{document}

\flushbottom
\maketitle

\thispagestyle{empty}

\section{Introduction}\label{sec1}

Anthropogenic climate change is an unprecedented challenge for humanity \cite{IPCC2021}, and the past decade has been the warmest in the last 125,000 years \cite{IPCC2022} with far-reaching consequences that influence both the natural environment and social systems \cite{schmidt2013media}. 

Climate change is a central topic in public debate \cite{david2022public}, environmental policy \cite{fekete2021review, change2001climate}, and public discourse. Tackling its catastrophic impacts within the time frame dictated by science is of paramount importance \cite{IPCC2018}. However, the process of transitioning away from fossil fuels, the main contributing factor to climate change, is hampered by an increasing polarised political debate \cite{falkenberg2022growing, birch2020political}. Therefore, understanding how the public perceives climate change and how this influences their actions is crucial for the effectiveness of climate policies.

Given its magnitude and the interdependence of the actors involved, climate change represents a multidimensional challenge: while it requires collective action and global coordination \cite{IPCC2014, change2014mitigation}, the process of transitioning away from fossil fuel falls heavily on sectors within people's private sphere of action (i.e. building and transports \cite{winkler2023effect}). In this context, improving public understanding of the phenomena, as well as the role that different social groups should play within the transition process, is crucial to transform the debate into concrete actions \cite{torricelli2023does} and achieving climate targets at regional and global levels.

Social media platforms play a crucial role in the dissemination of information related to climate change and climate policy. Previous research has explored the relationship between major events such as extreme climate events \cite{torricelli2023does, meyer2023between, mumenthaler2021impact, kirilenko2015people}, international conferences \cite{falkenberg2022growing}, and climate protests \cite{lochner2024climate}, and public interest in climate-related issues. These studies have generally found that public engagement with climate change increases when significant events occur, drawing widespread attention and prompting discussions. 

The topic of climate change is closely linked to the spread of conspiracy theories and misinformation \cite{candellone2024characteristics, treen2020online, van2017inoculating}, highlighting the need for effective countermeasures. These include fact-checking and corrections but should also extend beyond them by prioritizing the timely production of high-quality information \cite{cinelli2022promoting}. Ideally, this should occur before the emergence of so-called information voids, which serve as hotspots for misinformation \cite{aslett2024online, dataandsociety2019datavoids} due to the scarcity of relevant data and the lack of reliable information on topics of interest to online users \cite{purnat2021infodemic}.

In this paper, we tackle the relationship between information supply and demand within the climate change discourse in Italy, with a focus on their temporal responsiveness and degree of interconnection, in order to gain deeper insight into the dynamics of the information ecosystem. 
The analysis is carried out through a case study on selected climate-relevant topics (e.g. air quality, biofuel, electric cars). The considered topics were selected after consultation with experts working for an independent Italian non-profit think tank dedicated to climate change.
The research examines the supply and demand dynamics of climate change-related information on two major social media platforms, Facebook and Instagram. To date, the role of social media as an information source and a forum for public debate \cite{avalle2024persistent, aldayel2021stance, cinelli2021echo, tucker2018social} is well documented in the literature, including studies on climate change communication \cite{galdeman2025mapping ,pearce2019social, cody2015climate, schafer2012online}. 

Specifically, information supply is defined as the production of content on these platforms, with the analysis focusing on the Italian information ecosystem from December 2022 to August 2024. To complement social media data and enhance the robustness of the proposed analysis we introduce also an extra proxy of information supply related to news production, namely, the Global Database of Events, Language, and Tone (GDELT). Specifically, we use the GDELT Summary \cite{gdelt_summary_api} tool (see Methods for details), which aggregates and summarizes global news media coverage related to selected topics.

Information demand is modelled for the same geographical area and topics using Google Trends data. Google Trends provides insights into the relative popularity of search terms entered into the Google search engine, making it a widely accepted proxy for public attention, information demand, or topic salience \cite{timoneda2022spikes, holzl2025mis}. Its use is supported by both practical and methodological advantages. Practically, Google accounts for over 90\% of global Internet searches, ensuring broad and representative coverage. Methodologically, data are collected passively through direct observation of human behaviour \cite{mavragani2018assessing, googletrendsFAQ}, avoiding common survey biases such as recall errors or social desirability. 
 
We find that the demand of information (searches on Google Trends) often exceeds the supply of information (volumes of posts on social media and news articles). 
However, both dimensions fluctuate in tandem, especially in response to significant real-world events. Cross-correlation analysis confirms a temporal coupling, indicating that content production often responds promptly to shifts in public interest. These findings point to an overall adaptive information ecosystem, in which supply and demand dynamically align. This coupling also points to a form of feedback loop between content producers and information-seeking users. However, such responsiveness does not eliminate the risk of unmet demand, highlighting a critical problem: when users actively seek information that is not readily available, information voids may emerge.

To better understand these imbalances, we analyse the time series derived from the difference between supply and demand, enabling a dynamic characterization of both information voids (demand exceeding supply) and information overabundance (supply exceeding demand). Finally, we examine the relationship between these imbalances and social media engagement. While information voids are particularly pronounced in the context of the climate transition, our findings reveal no consistent correlation between engagement levels and the presence of either shortages or excesses of information. 

\section{Results}\label{sec2}

\subsection{Trends of supply and demand}\label{subsec0}
We study the information ecosystem as a market driven by two forces: supply and demand. Since these two quantities are likely to evolve over time and respond to both external events and mutual interactions, we begin by visualizing their trends over the period ranging from December 2022 to August 2024. To ensure comparability in terms of volume and time scale, we transformed Facebook, Instagram and GDELT data separately, aligning them with the scale and granularity of Google Trends. This rescaling step was essential, as search volumes, social media platforms activity and news production often differ substantially in magnitude (see Methods for details).

\begin{figure}[ht]
    \centering
    \includegraphics[width=0.9\linewidth]{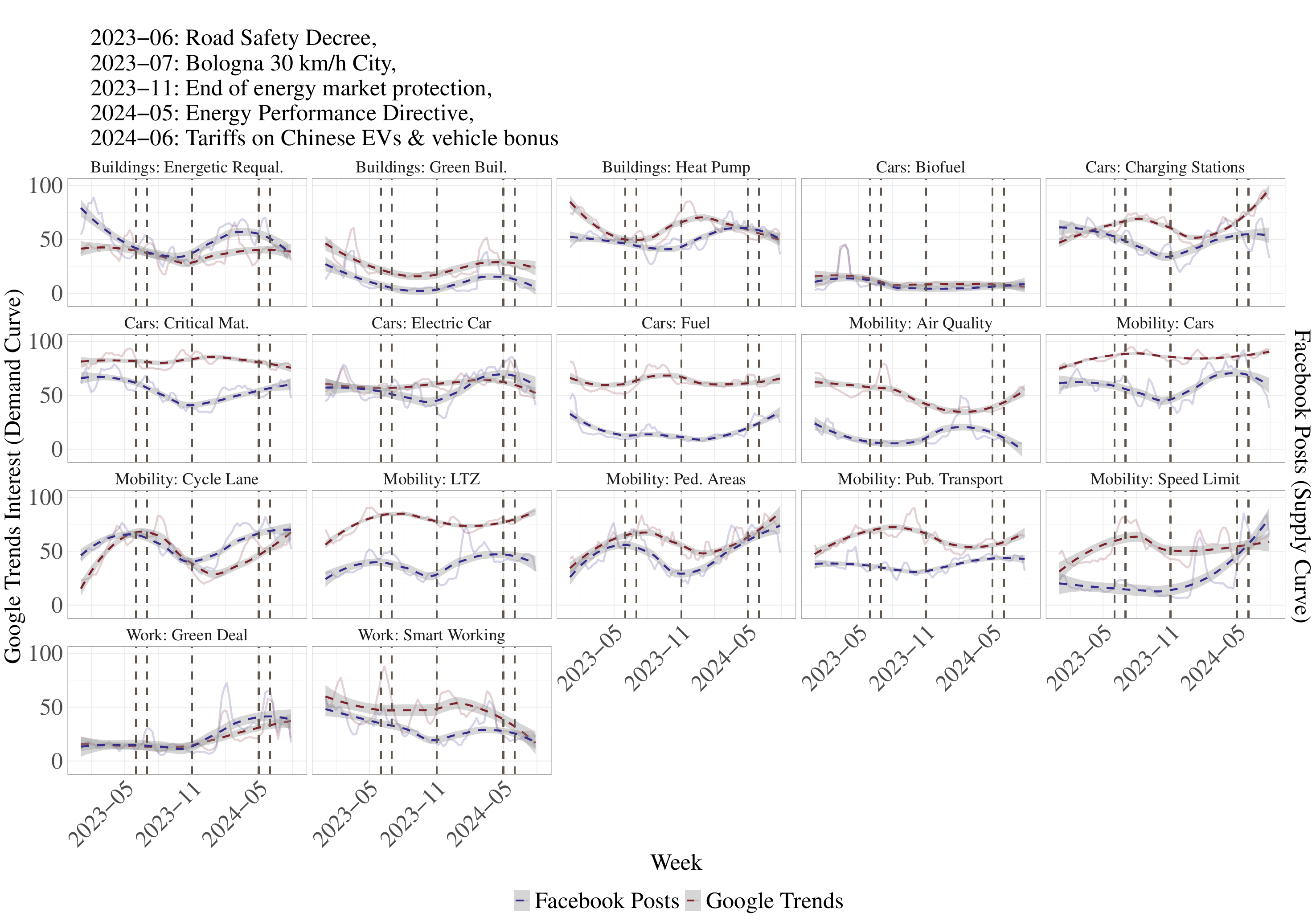}
  \caption{Time series for supply and demand of information. The data spans from December 2022 to August 2024, highlighting major events including the June 2023 Road Safety Decree, Bologna’s adoption of a 30 km/h speed limit in July 2023, the November 2023 end of energy market protection, the May 2024 Energy Performance of Buildings Directive (EPBD), and the June 2024 announcement of EU tariffs on Chinese electric vehicles and new vehicle bonuses. Dashed vertical lines indicate key policy or societal events, providing context for fluctuations in information demand and supply.}
   \label{fig:Supply_Demand_FB}
\end{figure}

Fig.~\ref{fig:Supply_Demand_FB} displays weekly time series for all the considered topics, one per each panel, on one of the two platforms - namely Facebook. The results for the other platform (Instagram) are reported in the Supplementary Information (SI) and show consistent findings. All results related to GDELT are also reported in the SI and confirm the overall dynamics observed on social media platforms.

We notice that demand (Google Trends) almost always locates above supply (Facebook posts in this case) and that the curves display similar shapes, suggesting a dynamic interaction between them. 
Furthermore, the curves appear to be sensitive with respect to external events, as observed particularly in the demand and supply curves for the topics \textit{Cycle Lanes} and \textit{Pedestrian Areas}. Notably, an increase in both demand and supply is evident during June and July 2023, coinciding with events such as the road safety decree and Bologna's decision to adopt the \textit{Città 30} (city 30 Km/h) model.

As the two curves are likely to display interesting reciprocal patterns we also performed a cross-correlation analysis in order understand if one of the two systematically precedes the other, that is if a greater supply of information is associated with a greater demand and vice versa. As reported in SI, the cross-correlation analysis confirms that, for the vast majority of topics, the highest correlation peak occurs at a temporal lag of zero, with the average of correlation values among the curves being around 0.35. This reflects a rapid responsiveness, at the considered time scale, of the information supply to variations in demand and vice versa.

\subsection{Differences between production and demand}\label{subsec1}
\begin{figure}[ht]
\centering
   \includegraphics[width=0.8\linewidth]{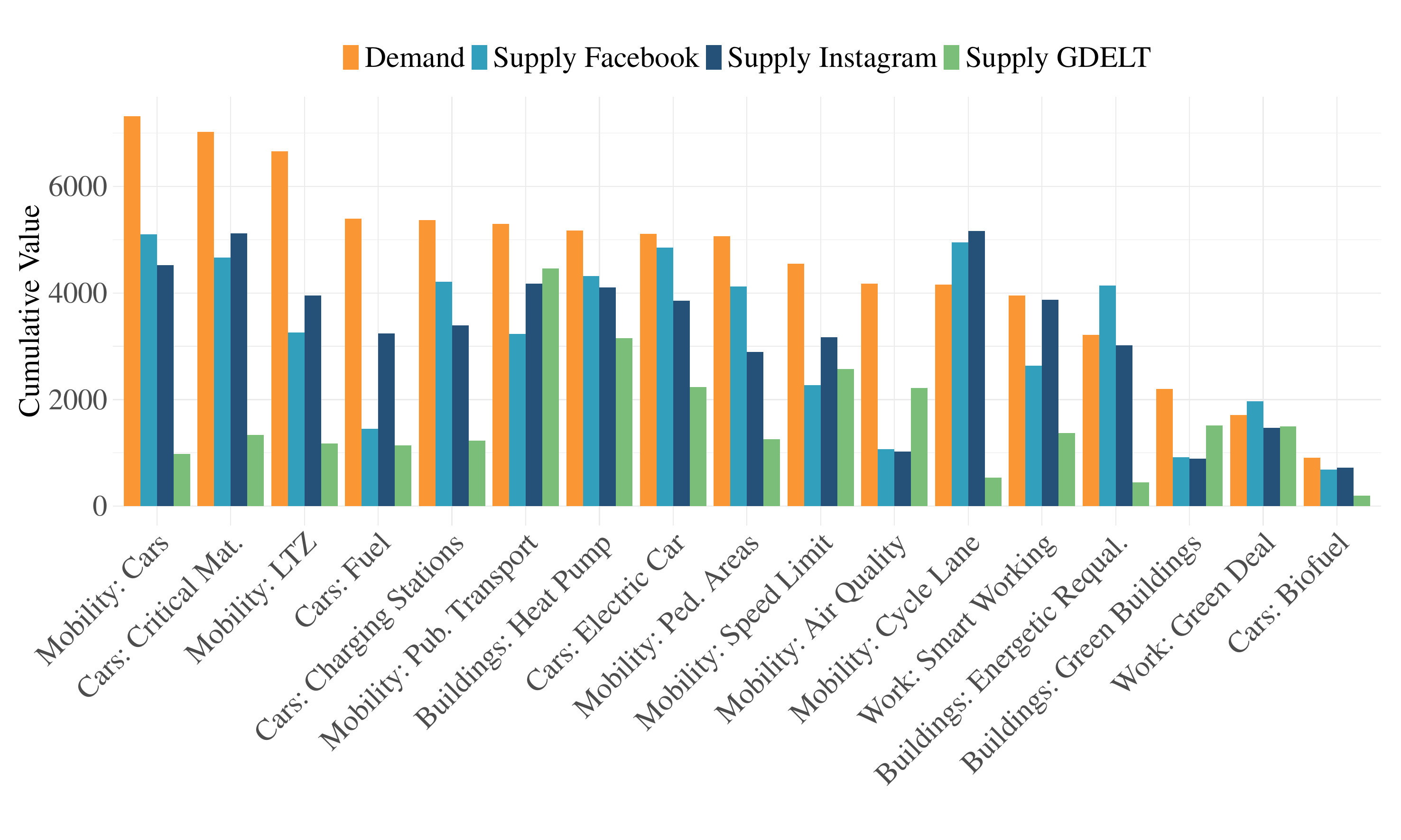}
  \caption{The cumulative count over the entire period (from December 2022 to August 2024) of the two forces, where Supply is reported separately for Facebook (FB), Instagram (IG) and GDELT across all keywords.}
   \label{fig:Cum_Value}
\end{figure}

After observing the trends of supply and demand we evaluate their overall values by visualising the cumulative sum of the two forces over the monitored period (see Fig.~\ref{fig:Cum_Value}). 
As observed, the demand for information consistently exceeds the supply from social media, regardless of the platform analysed. This phenomenon applies to all the considered topics, except for \textit{Cycle Lane}, \textit{Energetic Requalification}, and \textit{Green Deal}, where the supply of information surpasses demand. 
These phenomena have been characterized by significant media attention, driven by public interest and the political debates that have developed over the years under analysis. In particular, during the reference period, there has been a growing interest in the \textit{Green Deal} and \textit{Energetic Requalification}, largely due to the extensive discussions taking place both at the European Union and national level. At the same time, topics such as 
\textit{Cycle Lanes} have gained particular attention, partly as a result of the debate surrounding the \textit{Bologna Città 30} (city 30Km/h) initiative, which has gradually expanded to the national level.

Analysing news supply through GDELT reveals consistent patterns but significant differences in cumulative values across the various sub-topics: news supply is lower for \textit{Cars}, \textit{Critical Materials}, \textit{LZT}, \textit{Fuel}, and \textit{Charging Stations}, while it is higher for \textit{Public Transport}, \textit{Air Quality}, \textit{Green Buildings}, and \textit{Green Deal}. Despite the differing focus of news outlets compared to social media content, there is still an overall demand for information that exceeds the supply.

\subsection{Information excess and voids}\label{subsec2}
Building on the results of Figure 1 regarding the coupling between the supply and demand curves (also supported by the cross correlation analysis reported in SI), in this section we further examine the temporal alignment between supply and demand. To capture this dynamic, we introduce a metric called information delta, which allows us to assess potential excesses or voids in information supply relative to demand for a given topic.

In mathematical terms, the difference between supply and demand at week $w$ is:
\begin{equation}
    \delta_w = \hat{s}_w - \hat{d}_w
\end{equation}
where $\hat{s}_w$ is the normalized supply for social media posts and news articles as described in Eq.\ref{eq:norm_supply} and $\hat{d}_w$ is the normalized demand as provided by Google Trends (see Methods for details).

\begin{figure}[ht]
\centering
    \includegraphics[width=0.9\linewidth]{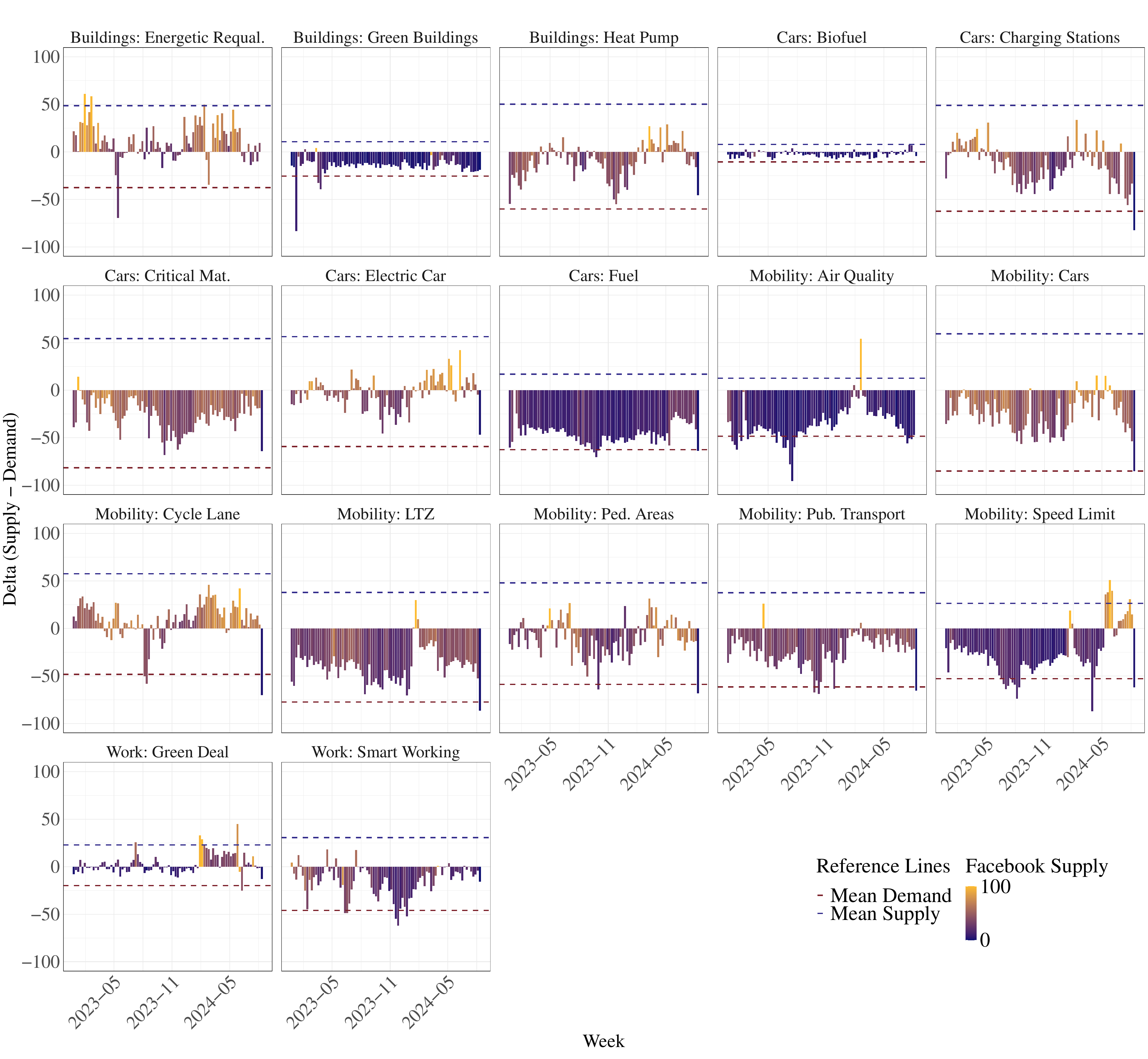}
  \caption{Weekly variation of the supply-demand delta across categories, with colour intensity representing Facebook supply levels. Each panel corresponds to a specific category, showing trends over time.}
   \label{fig:Delta_FB}
\end{figure}

This quantity ranges from -100 to 100 and positive delta values indicate an overabundance of information (supply exceeds demand) while negative values highlight information voids (demand exceeds supply). Overabundance and voids become better defined when their value surpasses two thresholds that we represent on Fig.~\ref{fig:Delta_FB}. In detail, these thresholds are represented by the average supply (above zero) and demand (below zero) for the considered topics. Thus, if the delta value surpasses the average supply for the whole time window it means that the discrepancy between supply and demand was even stronger than the average production value. The same reasoning holds in the case of information demand and voids.  
The visualization reveals distinct patterns across topics and time, with fluctuations in delta values that may reflect the dynamics of public interest and the responsiveness of content creators on social media. By incorporating the intensity of the Facebook supply (as indicated by the colour gradient), we can observe the relationship between the volume of content and the magnitude of the delta. This perspective helps to identify whether higher levels of supply reduce information voids or amplify overabundance. Among the sub-topics analysed, \textit{cars} exhibits the highest average information supply and demand values. In contrast, \textit{biofuel} records the lowest levels on both fronts highlighting a relatively stable relationship between supply and demand.

Across all topics, a consistent and persistent lack of information is observed, suggesting that content supply often falls short of public demand. Notably, exceptions include \textit{Cycle Lane}, \textit{Green Deal}, and \textit{Energetic Requalification}, which display longer periods of abundance of information. The equivalent visualizations of delta values for Instagram and GDELT are provided in SI.

\subsection{Effect on engagement}\label{subsec3}
\begin{figure}[ht]
    \centering
    \includegraphics[width=0.9\linewidth]{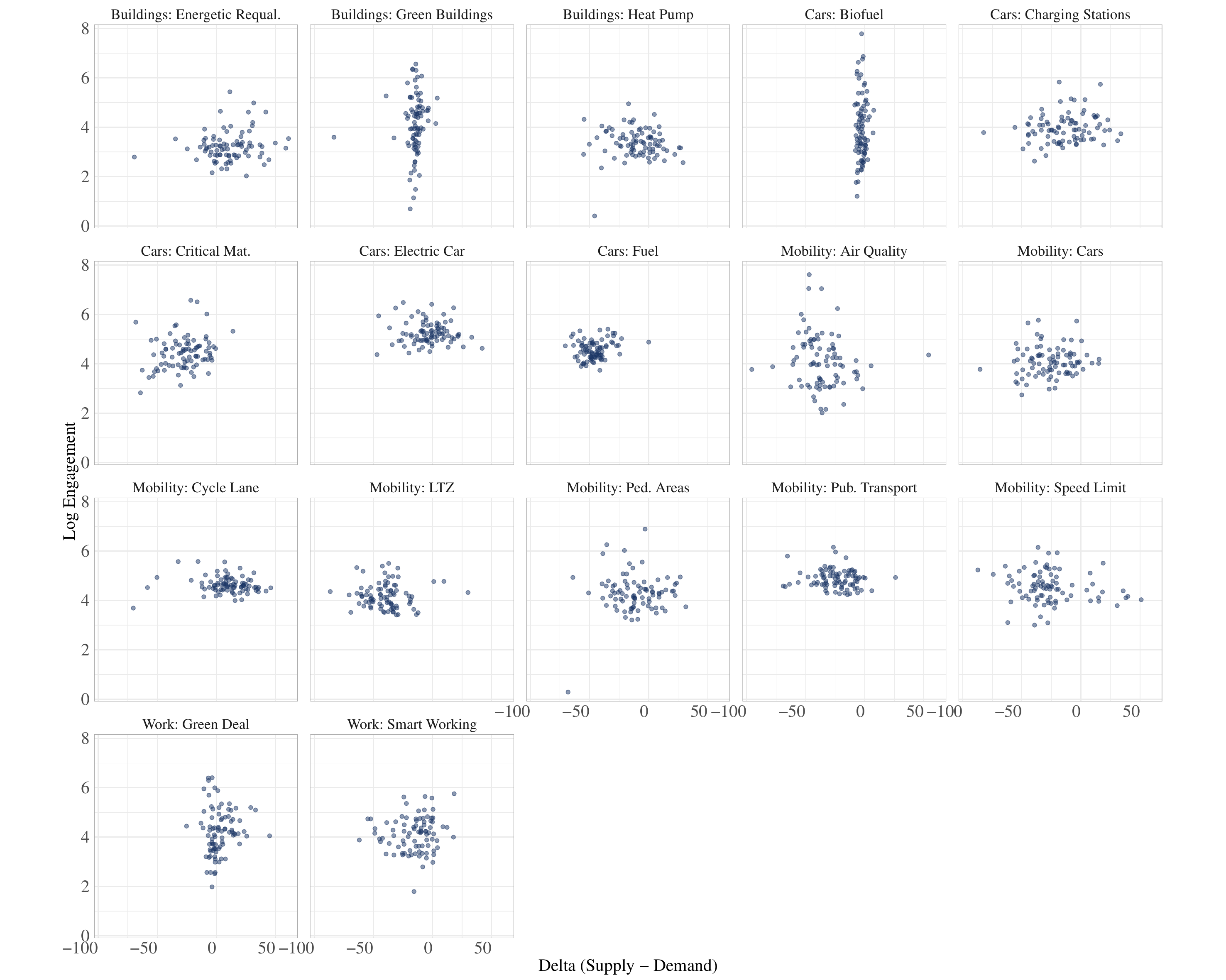}
    \caption{Relationship between supply-demand gap and engagement, log-transformed engagement values across different categories.}
    \label{fig:reg_fb}
\end{figure}

In order to understand the sensitivity of users response to demand (Google Trends) and supply (in this represented only by the volume of social media posts), we analyse the relationship between the delta values (the gap between information supply and demand) and user engagement, measured as the total number of interactions (e.g. likes) with posts. Our findings, shown in Fig. \ref{fig:reg_fb}, indicate that the information delta does not display significant correlation with engagement. 
Indeed, the scatter plots show mostly high dispersion and thus a lack of association (in either direction between the variables). This is further supported by the average correlation between engagement and delta across all categories, which is -0.04 for Facebook and -0.02 for Instagram, reinforcing the absence of a clear relationship. Upon a more detailed scrutiny, performed by means of a panel regression model that includes the number of followers held by the considered accounts and reported in SI, we note that users' engagement appears to be more strongly associated with the number of followers at the time of posting. This suggests that user engagement behaviour is not strictly dependent on the information delta, but rather on the audience size of the page. Furthermore, attention to climate transition-related topics seems to be influenced by external events, highlighting how public interest can be driven by broader contextual factors rather than just the available information.

\section{Conclusions}\label{sec4}

Information quality is a crucial challenge in the Information Age, particularly in the context of the ongoing climate crisis and the urgent need for a successful climate transition. The transformations outlined in the IPCC report require more than just technological advancements or improved climate models—they depend on an informed, engaged, and empowered society, able to navigate complex and highly technical issues \cite{IPCC2022, devine2022placing}. Addressing the persistent information voids surrounding the climate transition is essential to fostering a more responsible and proactive public, that more and more is expressing the need for reliable and easy to access information.  Without accessible, accurate and timely information, people can struggle to make informed decisions, ultimately preventing their participation in designing and implementing collective solutions to climate change \cite{ECCO2024}.  

The dynamics of information supply and demand on climate-related topics tend to intensify in response to significant external events, underscoring how public attention and social media discourse are shaped by these occurrences. Rather than being driven by information voids, engagement appears to be more closely linked to moments of heightened media relevance. Notably, spikes in both information production and demand coincide with regional policy initiatives, such as \textit{Bologna Città 30}, and broader climate transition measures at the European level. This pattern becomes even more pronounced during major international events, such as the COPs \cite{hase2021climate, falkenberg2022growing, schafer2014drives}, or in response to extreme climate phenomena.


This article reveals that information supply and demand on climate change and climate transition topics are not only interrelated but also exhibit clear temporal coupling. Notably, supply tends to respond promptly to spikes in public interest, particularly during moments shaped by significant external events. These patterns suggest the presence of an overall adaptive information ecosystem, in which content producers react dynamically to changing information needs. However, despite this responsiveness, demand consistently exceeds supply over the observed period, pointing to persistent and recurring information voids.

This structural mismatch implies that users seeking climate-related content may struggle to find timely or adequate information, potentially hindering public understanding and delaying meaningful engagement. Given the central role of news outlets and social media as both an information source and a forum for public discourse, such gaps raise important concerns about the accessibility and relevance of climate-related content. Furthermore, although supply and demand generally follow similar trajectories, we find no consistent correlation between user engagement and the presence of either information voids or excesses. This may, in part, be due to the limited temporal granularity of Google Trends, which provides data only at a weekly scale. A more detailed, daily-level analysis—requiring a different proxy for demand—could help clarify the relationship between engagement and information imbalances. It is also important to emphasize that Facebook and Instagram users are not fully representative of the general public in terms of demographics, interests, and behaviours. Information on climate change and climate transition is also disseminated through a wide range of other channels, including traditional media and alternative online platforms. The GDELT analysis also shows an information supply deficit, confirming that this imbalance is not limited to social media platforms alone.

Future research should further refine the methodological framework, improving its analytical rigour while extending the investigation to other digital platforms in order to examine the dissemination of information and the nature of public debate on issues related to climate change and climate transition. 

In summary, this research introduces an innovative approach for analysing and quantifying the volume and dynamics of information on climate change-related topics, with particular attention to the responsiveness and flexibility of the information ecosystem, as well as the persistence of information voids. The proposed theoretical framework is designed to be both adaptable and transferable, making it suitable for application across diverse thematic and geographical contexts. By mapping the interplay between information supply and demand, this approach provides a valuable foundation for broader investigations into public discourse and media responsiveness. Finally, identifying information voids within the climate debate presents an opportunity to increase consensus at national level around key decarbonization policies and ensuring, as a consequence, that the global community stays on track to reach the climate targets collectively agreed.

\section{Methods}\label{sec6}

\subsection{Topic Selection}\label{subsec6}
This section outlines the procedures employed for data collection and preprocessing and provides detailed information about the datasets used in this study. The topics were selected through collaboration and consultation with a group of experts from an independent, non-profit Italian think tank dedicated to climate change.
The analysis covers four primary thematic areas (topics) that are relevant in the context of the climate transition, namely: Buildings, Cars, Mobility, and Work. In fact, building and transport together account for 58\% of global energy consumption and 26\% of global emission \cite{van2025demand}, while addressing the social challenges of the transition for workers is essential to ensure that everyone benefits from the opportunities created by the process itself. For each category, different subtopics were identified as follows. In the case of Buildings the subtopics are: Green Buildings, Energetic Requalification, Heat pumps. In the case of Cars the subtopics are: Electric Car, Charging Stations, Fuel, Biofuel, Critical Materials. In the case of Mobility, the subtopics are: Cars, Public Transportation, LTZ, Air Quality, Speed Limits, Pedestrian Area, Cycle Lane. In the case of Work, the subtopics are: Fair Transition, Green Deal, Smart Working.  

\subsection{Data Collection}\label{subsec7}
The research focuses on analysing the interplay between two fundamental forces: supply and demand of information, over the period ranging from December 2022 to August 2024. According to previous work \cite{gravino2022supply,gravino2024online}, we used posts on Facebook and Instagram on the aforementioned topics as a proxy for information production and data from web searches, quantified by Google Trends, as a proxy for information demand. As a third proxy for information supply, we also incorporated data from GDELT, which captures global news media coverage. All details regarding data sources, collection procedures, and preprocessing steps are provided in the SI.

\subsection{Social Media Data}\label{subsec8}
Social media data were collected from CrowdTangle, a Meta owned tool for data collection that was active until August 14 2024. CrowdTangle provided access to a limited set of accounts. For Facebook, the dataset included over 7 million public Facebook pages and verified profiles. Specifically, it comprised all public Facebook pages with more than 50,000 likes, all public groups with over 95,000 members, all US-based public groups with more than 2,000 members, and all verified profiles. 

For Instagram, the data set included more than 2 million public Instagram accounts, consisting of all public accounts with more than 50,000 followers and all verified accounts.
To gather the data we employed the keyword search function providing as input relevant keywords for the search. For instance, in the case of the topic Buildings, sub-topic Green Buildings we used the following query: \textit{``casa green'' OR  ``case green'' OR ``EPBD''} where ``casa green'' means ``green building'', ``case green'' means ``green buildings'' and EPBD refers to Energy Performance of Buildings Directive, i.e. a legislative framework that establishes a range of measures to improve the energy efficiency of buildings across Europe. The full list of strings used for the search is reported in SI.

After an initial data collection phase, we distinguished between two types of Facebook content: posts published on public pages or user profiles, and posts shared within Facebook groups. We observed that posts from groups often introduced substantial noise into the analysis. For instance, in the sub-topic \textit{``mobility\_public\_transport''}, the most frequently used hashtags included terms such as \#affitto (means \textit{``rent''}), \#cerco (means \textit{``looking for''}), \#offro (means \textit{``offering''}), and \#studenti (means \textit{``students''}). These hashtags were largely unrelated to the scope of the study and were commonly found in private groups used for housing or student life. Such content significantly skewed the semantic and topical relevance of the dataset. To address this, we applied a filtering step to exclude all posts originating from Facebook groups. This allowed us to retain only content that more reliably reflected public discourse rather than community-specific or transactional conversations. Following this cleaning process, the final dataset includes 412,332 Facebook posts from 34,921 unique accounts and 45,122 Instagram posts from 9,095 unique accounts, covering the period from December 2022 to August 2024. CrowdTangle also provides engagement for the posts that, in the case of Facebook, corresponds to Likes, Comments, Shares and Reactions while in the case of Instagram to Likes and Comments. For each post CrowdTangle also provides the Total Engagement metric which is the sum of the previously mentioned interactions for each post. The grand sum of total engagement in the case of Facebook data is 49,147,635 while in the case of Instagram is 57,652,139. Table ~\ref{tab:tab1} provides a detailed breakdown of the number of posts, unique accounts and total engagement by platforms, offering an overview of the dataset's structure and scale. 

\begin{table}[ht]
    \centering
    \begin{tabular}{l|l|l}
        \hline
        & \textbf{Facebook} & \textbf{Instagram} \\
        \hline
        \ Time window from  & 2022/12/26 & 2022/12/26 \\
        \ Time window to  &  2024/08/14 & 2024/08/14 \\
        \ Number of posts & 412,332 & 45,122 \\
        \ Unique Accounts & 34,921 & 9,095 \\
        \ Total Engagement & 49,147,635 & 57,652,139 \\
        \hline
    \end{tabular}
    \caption{Data breakdown for Facebook and Instagram posts.}
    \label{tab:tab1}
\end{table}

\subsection{News data}\label{subsec8.1}
\begin{figure}[ht]
\centering
   \includegraphics[width=0.6\linewidth]{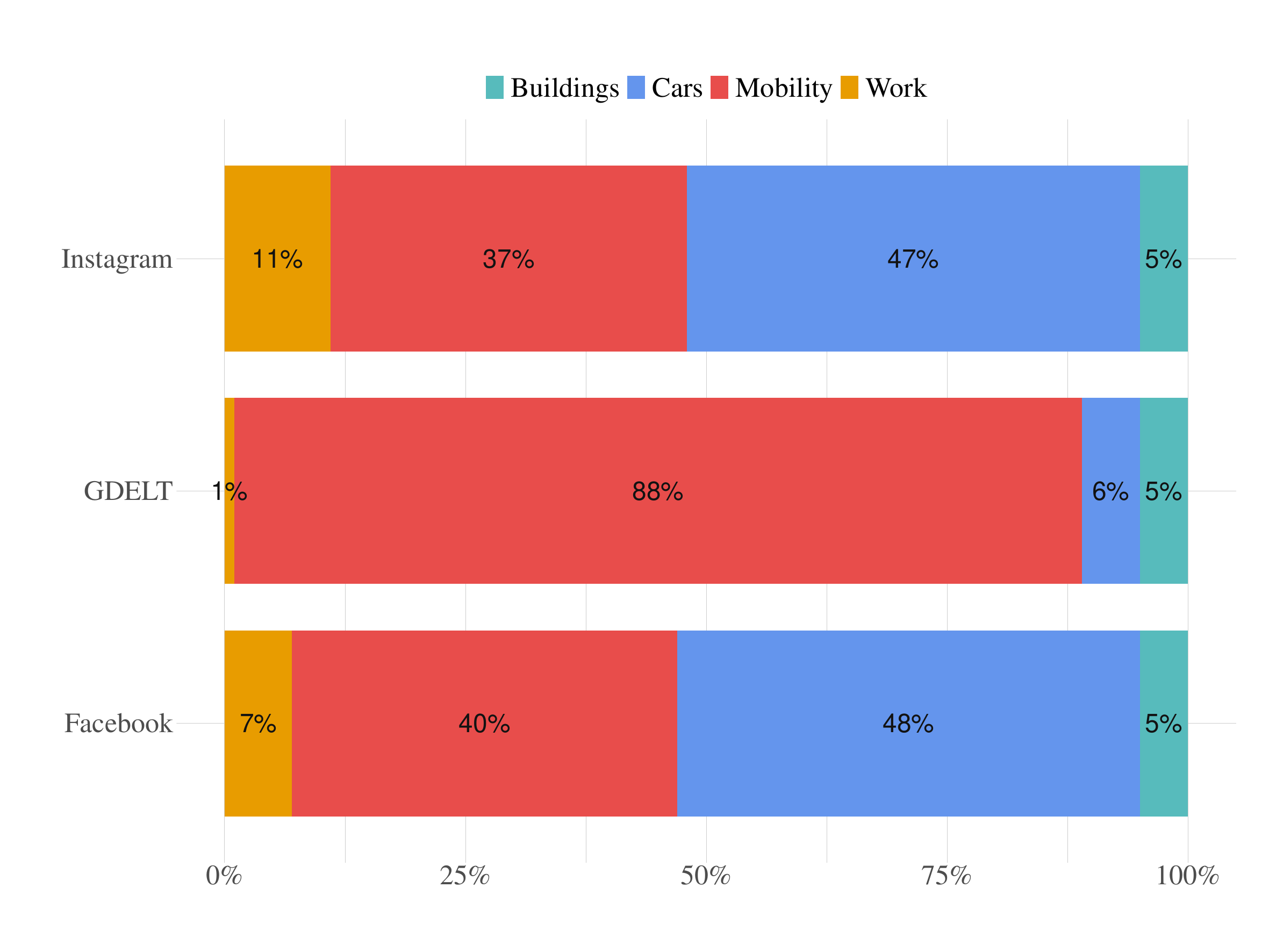}
  \caption{Comparison of category distributions between Facebook, Instagram and GDELT. The stacked bars represent the percentage composition of content related to four macro categories: Buildings, Cars, Mobility, and Work on each platform.}
   \label{fig:Val_percent}
\end{figure}

GDELT data were collected using the GDELT Summary platform. This platform enables the creation of customized dashboards that offer a synthesized overview of global news coverage including print, online and local news. For this study, a keyword-based search was performed, focusing on the \textit{``Global Online News Coverage''} stream. The keywords used for data extraction align with those employed for collecting data from Facebook and Instagram (the complete list is provided in SI). The search keywords must be entered in English, as GDELT automatically translates non-English news content into English for indexing and search purposes, with the option to filter by country—in this case, Italy. Additionally, the platform allows users to review the articles within the time series output, ensuring the robustness and reliability of the results. The dataset includes 24,004 online articles covering the period from December 2022 to August 2024. Figure~\ref{fig:Val_percent} shows the distribution of posts and news articles by topic and source.
\subsection{Search data}\label{subsec9}
Among non-traditional data sources \cite{cebrian2023google}, Google Trends is one of the most widely used in the socio-economic literature \cite{costola2021google, preis2013quantifying} and has proven to be a valuable resource for analysing human behaviour in various contexts \cite{jun2018ten, zhang2018using, choi2012predicting}.
Google Trends offers access to anonymized data based on actual search queries made on the Google Search Engine, providing insights into the frequency of specific search terms across different locations and time periods. For the purposes of this study, search data was obtained for Italy, spanning from December 26, 2022, to August 12, 2024. 
Google Trends allows users to download data through topic-based searches, which include a set of terms sharing the same concept, regardless of the language. For example, searching for \textit{``London''} and selecting the corresponding topic will yield results related to concepts such as \textit{``Capital of the United Kingdom''} and \textit{``Londres''}, the Spanish term for \textit{``London''}. In this analysis, data were primarily collected using the topic-based search feature, with the exception of three subtopics: \textit{``Casa Green'', ``Riqualificazione Energetica'', ``Area Pedonale''} where ``Casa Green'' means ``green building'', ``Riqualificazione Energetica'' means ``Energetic Requalification'' and ``Area Pedonale'' means ``Pedestrian Aerea''. For these subtopics, data were obtained using keyword search, as no results were available through topic-based searches. A comprehensive list of the search terms can be found in the SI. 
Regardless of the search method used, Google Trends provides the data as weekly time series, with each value representing search interest relative to the highest point within the selected period. All time series, whether topic-based or keyword-based, are normalized on a scale from 0 to 100, where 100 represents the highest recorded search volume within the selected time frame. A value of 0 does not necessarily indicate the absence of searches but rather denotes volumes too low to be reported. The normalization process also excludes queries made within a short time frame from the same IP address and those containing special characters \cite{mavragani2019google}. This allows the retrieval of normalized search interest for any keyword, regardless of additional filters. As Google does not disclose absolute search counts, the reported values represent percentages of the peak search volume within the analysed period. Consequently, data consist of integers from 0 to 100, where 50 indicates half of the maximum observed searches.

\subsection{Data aggregation}\label{subsec10}
In order to make social media, news and search data comparable in terms of volume and time scale we performed a transformation of Facebook Instagram and GDELT data separately in order to consistently match the scale and granularity retrieved from Google Trends. In detail, we aggregated social media data at a weekly time scale by adding up all the posts produced during the week and then rescaled the volume of posts in order to fit within the 0-100 scale where 100 corresponds to the maximum number of social media posts/news per each category in the considered time window. 

In mathematical terms, the re-scaled value of the number of weekly posts $s_w$ on a given platform is:
\begin{equation}
    \hat{s}_w = \biggl\lfloor \frac{s_w}{ \max \limits_{w \in W} s_w} + 0.5 \biggl\rfloor
    \label{eq:norm_supply}
\end{equation}
where $W$ is the set of weeks in the time window ranging from 2022/12/26 to 2024/08/14.

\section*{Declarations}

Giulia Colafrancesco is employed by ECCO non-profit climate change think tank.

\end{document}